\documentstyle[aps,epsfig]{revtex}
\begin{document}
\title{Criticality of natural absorbing states}
\author{Adam  Lipowski$^{1)}$\cite{byline1} and Michel
Droz$^{2)}$\cite{byline2}}
\address{$^{1)}$ Department of Physics, A. Mickiewicz University,61-614 Poznan, 
Poland\\
$^{2)}$ Department of Physics, University of Gen\`eve, CH 1211
Gen\`eve 4, Switzerland}
\date{\today}
\maketitle
%%%%%%%%%%%%%%%%%%%%%%%%%%%%%%%%%%%%%%%%%%%%%%%%%%%%%%%%%%%%%%%%%
\begin{abstract}
We study a recently introduced ladder model which
undergoes a transition between an active and an infinitely degenerate
absorbing phase.
In some cases the critical behaviour of the model is the same as that of the
branching annihilating random walk with $N\geq 2$ species both with and
without hard-core interaction.
We show that certain static characteristics of the so-called natural
absorbing states develop power law singularities which signal the approach
of the critical point.
These results are also explained using random walk arguments.
In addition to that we show that when dynamics of our model is
considered as a minimum finding procedure, it has the best efficiency very
close to the critical point.
\end{abstract}
%%%%%%%%%%%%%%%%%%%%%%%%%%%%%%%%%%%%%%%%%%%%%%%%%%%%%%%%%%%%%%%%%%%%%%%%%%%%%%
\section{Introduction}

In recent years, a large body of theoretical works~\cite{HAYE1} has been
devoted to the study of phase transitions in model with absorbing states. One
aim of these studies is to characterize the possible universality classes.
Only  few exact results can be obtained. Accordingly, numerical simulations,
particularly Monte-Carlo simulations and finite size scaling methods, have
been widely used~\cite{AUKRUST}.

Usually, Monte Carlo method is used to simulate a model in its active phase,
where various steady-state characteristics can be measured.
When the system approaches a critical point these characteristics develop some
critical singularities.
But the critical behaviour can be also studied  from an absorbing
phase.
In this case, however, one usually studies certain dynamical properites of the
model which exhibit critical singularities.
For example, the average time needed to reach an absorbing state diverges
when  approaching the critical point.
When appropriately defined, all critical exponents can be defined through some
dynamical singularities in an absorbing phase.

Is it possible to infer some information about criticality of the
model from the structure of absorbing state(s)?
Very often a given model has only finitely degenerate, homogeneous absorbing
state which do not carry any of such information.
However, there exists a number of models which have infinitely many
absorbing states.  When the control parameter(s) set the model in the
absorbing phase and far from the critical point, an absorbing state is
reached quite fast and
we might expect that, in such a case, the absorbing state will be
almost randomly selected among all the absorbing states of the model. 
In the following, an absorbing state reached as a consequence of the dynamics
of the model and starting from a random initial configuration will be called a
natural absorbing state. 
However,  when the system is close to the critical point, the evolution
towards the absorbing state is quite long and can be complicated.  
It is very likely that, in this case, natural absorbing states constitute
only a certain fraction of all absorbing states. 

Are there any quantities which would specify which absorbing states are
selected by the model's dynamics?  
In the context of phase transitions
it would be particularly interesting if such quantities would also
contain information about the critical point of the model (e.g., its
location and maybe some of its exponents).
We are not aware that, such questions, which would certainly provide a better
understanding of models with absorbing states, have been previously considered.
 
The aim of the present paper is to investigate the above questions by
studying a recently introduced one-dimensional model with infinitely
many absorbing states~\cite{LIP2001}.
Critical behaviour of this model seems to be closely related with that of some
multispecies branching-annihilating random walk models (BARW).
Using Monte Carlo method we show that the natural absorbing states  do
contain important information about the critical behaviour of the
model.  
In particular, we show that certain characteristics of natural
absorbing states exhibit power singularities upon approaching the
critical point.  

The model and some of its basic properties are
described in section 2.
Monte Carlo results which show that the structure of natural absorbing
states contains information about criticality of the model are discussed in
section 3.  
In section 4, numerical findings are corroborated by
random-walk arguments.  
In section 5 we also look at the dynamics of our model as a certain
minimum-finding procedure.
Such an approach, when suitably generalized might serve as a
general-purpose technique to solve some optimization problems.
Finally, concluding remarks are made in section 6.
%%%%%%%%%%%%%%%%%%%%%%%%%%%%%%%%%%%%%%%%%%%%%%%%%%%%%%%%%%%%%%%%%%%%%%%%
\section{Model and its basic properties}

Our model is defined on a one-dimensional ladder-like lattice.  For
each bond between the nearest-neighbouring sites we introduce a bond
variable $w\in (-1,1)$~\cite{COMM1}.  Introducing parameters $r$ and
$s$, we call a given site active when $w_1w_2|w_3|^s<r$, where $w_1$
and $w_2$ are intra-chain bond variables connected with this site and
$w_3$ is the inter-chain variable.  Otherwise, this site is called
nonactive.  The model is driven by random sequential dynamics and when
the active site is selected, we assign anew, with uniform probability,
three neighbouring bond variables.  Nonactive sites are not updated.
For a more detailed description of this model see~\cite{LIP2001}.

Models of this kind might be used to describe coarse-grained biological
evolution, where species (sites) mutate only when the interaction with other
species puts to much pressure on it (i.e., when the site is
active)~\cite{LIP2000}.
In addition, the above model provides an example of a coupled system whose
criticality is changed due to this coupling.
Recently, related systems are drawing certain attention~\cite{GEZA2001}.

Monte Carlo simulations supported by some analytic considerations
clarified the basic properties of this model, which can be briefly
sumarized as follows~\cite{LIP2001}. For $r>0$ the model is in the
active phase.  For $r<0$ the model generates, at a finite rate, sites
which are nonactive and which will remain in this state forever.  Thus,
the model, very quickly, enters an absorbing state and for $r<0$ the
model is in the absorbing phase.
The transition point $r=0$ is  critical and some quantities exhibit a power law
behavior. 
For example, the order parameter $\rho$ decays in time with an exponent
$\delta$ 
\begin{equation} \rho(t) \sim t^{-\delta} \ \ \ {\rm with} \
\delta=0.5 \label{2a}
\end{equation}
The average time $\tau$ needed to reach an absorbing state increases with the
system size $N$ as:  
\begin{equation}
\tau \sim N^z, \ \ \ {\rm with} \ z=2
\label{2b}
\end{equation}
Moreover, for $r \to 0^+$, the order parameter scales as $\rho \sim
r^{\beta}$, while the transverse   correlation length diverges as $\xi_{\perp}
\sim r^{-\nu_{\perp}}$  with: 
\begin{equation} \beta=\nu_{\perp}=1/s \ ({\rm for} \ s\geq 1). 
\label{3} 
\end{equation}

At $r=0$ the model's dynamics is particularly simple since in this case the
coupling between chains ($w_3$) is irrelevant and only the sign of $w_1w_2$
determines (non-)activity of a site.
The dynamics of the model is thus similar to a certain branching-annihilating
random walk model, for which exponents (\ref{2a})-(\ref{2b}) can be derived
analytically~\cite{AVRAHAM}.

At first sight, from (\ref{3}) it seems that our model has continuously
changing (with $s$) exponents $\beta$ and $\nu_{\perp}$.
However, it was shown~\cite{LIP2001} that this nonuniversality can be removed
when the control parameter $r$ is replaced by the reactivation
probability~\cite{REACT}.
Note that the critical behaviour of this model is
related with that of certain branching-annihilating random walk (BARW) models.
Indeed, critical exponents (\ref{3}) for $s=1$ are the same as in the
multispecies BARW model without exclusion~\cite{TAUBER}. 
Moreover, exponents for $s=2$ correspond to the BARW model with hard-core
exclusion~\cite{KWON,GEZA}.
Since nonuniversality of our model can be removed by an appropriate
redefinition of the control parameter, it suggests that a similar situation
might take place in BARW models.
However,  up to now it seems, that there is no simple relation between BARW
models with and without exclusion. 
In addition, it would be interesting to generalize BARW models so that
exponents $\beta$ and $\nu_{\perp}$ would change continuously with a certain
parameter which would interpolate between the case with and without exclusion.
%%%%%%%%%%%%%%%%%%%%%%%%%%%%%%%%%%%%%%%%%%%%%%%%%%%%%%%%%%%%%%%%%%%%%%%%%%%
\section{Structure of natural absorbing states}
In the present section we
examine natural absorbing states of the model for $r\leq 0$. First, let us
define the following pseudo-energy $e$ as:  
\begin{equation}
e=-\frac{1}{2N}\sum_j w_{1j}w_{2j}|w_{3j}|^s,  
\label{1}  
\end{equation} 
where
summation is over all sites $j$ of the model, $w_{1j},\ w_{2j}$ and $w_{3j}$ are three bond
variables connected to the site $j$ and $N$ is the linear size of the system.
Active sites are those whose contributions to~(\ref{1}) exceeds $-r$. 
Another  interesting quantity is the average value $\bar{e}(r)$ of $e$ calculated
for natural absorbing states. 

Note that at criticality, $\bar{e}(0)$ can be calculated exactly.
Indeed, in this case any of the absorbing states has the following
structure: each chain has all intra-chain variables $w$ of the same
sign and inter-chain variables can take arbitrary values.  Moreover,
for $r=0$ the dynamics of the model is such that it is only the sign of
bond variables which matters while their values are irrelevant.  It
means that the only requirement for a site belonging to a certain
absorbing state is that two intra-chain variables connected to this
site are of the same sign. Hence:
 
\begin{equation} 
\bar{e}(0)=-\frac{1}{2N}(2N)\int_0^{1}w_1dw_1\int_0^{1}w_2dw_2
\int_{-1}^{1}\frac{1}{2}|w_3|^sdw_3=-\frac{1}{4(s+1)}, 
\label{4}
\end{equation} 
where the triple integral gives the single site contribution to
$\bar{e}(0)$.

In the absorbing phase, the situation is not so simple and an analytic evaluation of $\bar{e}(r)$ is not possible. However, this quantity can be estimated by Monte-Carlo simulations. For a given value of $r$, we start from a random bonds configuration and evolve the system until an
absorbing state is reached. 
The average $\bar{e(}r)$ is then obtained by calculating $e$ for several
independent absorbing states.

Simulations were done for systems of size $N=1000$. 
For $r=0$, the analytical prediction was recovered with a great precision by 
averaging over approximately 
$10^3$ independent absorbing states.  
As a check, the result of (\ref{3}) for $s=1,2$ and 4 were reproduced with 
the accuracy of $10^{-3} \sim 10^{-4}$.

\begin{figure}
\epsfig{file=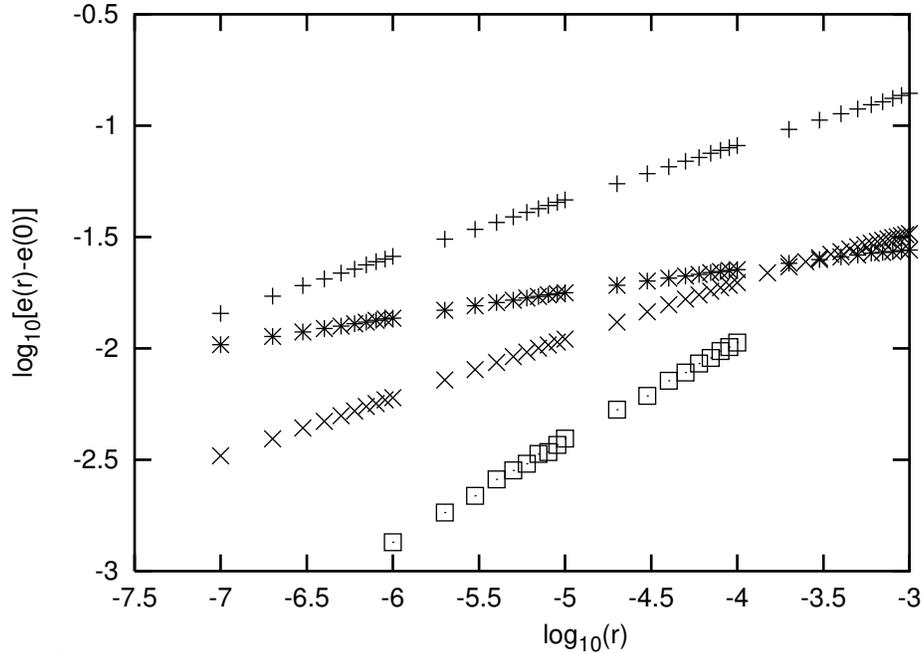, angle=-90}
\caption{ The log-log plot of $\bar{e}(r)-\bar{e}(0)$ as a function of $r$
for $s=1$ ($\Box$), 2 ($\times$) and 4 ($\star$). 
For $s=2$ we also plot the density of  permanently nonactive sites (+).}  
\label{f1}
\end{figure}

Our main concern is, however, the off-critical behaviour of $\bar{e}(r)$.
For $r<0$ the requirement that all intra-chain bonds have to be of the same sign is no longer necessary.  As a result, absorbing states might have a more
complicated structure, hence we were not able to calculate $\bar{e}(r)$ exactly. However, since for $r<0$ the dynamics leads to an absorbing state  much faster than
at criticality, we expect that $\bar{e}(r<0)>\bar{e}(0)$ and that $\bar{e}(r)$ is a decreasing function of $r$. Indeed, the dynamics searches for states with small pseudo-energy. This behavior is confirmed by the numerical simulations. 
In addition, as shown  in Fig.\ref{f1}, our results indicate that $\bar{e}(r)$ has a power-law singularity at $r=0$ and for vanishing $r$ behaves as 
\begin{equation}
\bar{e}(r)-\bar{e}(0)\sim r^p,
\label{5}
\end{equation}
where $p$ is an $s$-dependent exponent.

The least square fit to the these data (using the ten closest datapoints to $r=0$) leads to the following exponents: $p=0.46, 0.245$ and  0.128 for
$s=1,2$ and 4, respectively. These values of $p$ suggest that the true exponents (for infinite systems) are $p=\frac{1}{2s}$.
In the next section we present a random-walk  argument which supports 
this claim.

%%%%%%%%%%%%%%%%%%%%%%%%%%%%%%%%%%%%%%%%%%%%%%%%%%%%%%%%%%%%%%%%%%%%%%%%%%%
\section{Random walk argument}   First let us recall certain properties of
this model which holds for $r<0$. As it was already shown~\cite{LIP2001} in
this case the model generates sites which remain permanently nonactive.
Indeed, if after updating  certain active site the interchain bond $w_3$
satisfies the condition  \begin{equation} |w_3|^s<-r  \label{6}
\end{equation}
then both sites connected to $w_3$ remain permanently nonactive.
Since there is a finite probability to satisfy (\ref{5}), for $r<0$ activity in
the system quickly dies out and the model is in the absorbing phase.

Now, to explain the observed scaling of $\bar{e}(r)-\bar{e}(0)$,  let us
assume that $r$ is negative and very close to zero.
As it was already discussed, at $r=0$ an absorbing state is composed of
'ferromagnetic' chains.
We think that the following scenario describes the dynamics for $r<0$. 
For early times, the evolution of the model resembles the $r=0$ case and the system develops larger and larger 'ferromagnetic' domains (coarsening).
At the late stage the evolution of the system is mainly governed by the dynamics of the domain walls.
For $r=0$ the system would coarsen until the fully 'ferromagentic' state was reached.  However, for $r<0$, the activity between domains might die out quicker, mainly due to the possibility of creation of
permanently absorbing sites (in the following we will present some Monte Carlo
data supporting this assumption).

Using this assumption we estimate the number of permanently nonactive sites $N_p$ as follows.
First, let us notice that for $r=0$, the  coarsening proceeds with the exponent
$\delta=0.5$~\cite{LIP2001}. It means that the number of active sites $N_a(t)$ in the system scales for large time as $N_a(t) \sim Nt^{-0.5}$.
It is well known that the time needed to form a domain of size $l$ by coarsening (and for a dynamics with non conservation law) scales as $l^2$
(random walk argument~\cite{BRAY}).
Next, we estimate the total number of updates needed to create domains of the
typical size $l$ as:
\begin{equation}
\int_0^{l^2} N_a(t)dt\sim Nl.
\label{7}
\end{equation}
From (\ref{7}) the probability of creating of a permanently nonactive site scales as 
$(-r)^{1/s}$.
It means that  the typical size $l$ which scales as the inverse of the density
of permanently nonactive sites behaves as:
\begin{equation}
l \sim \frac{N}{Nl|r|^{1/s}}.
\label{8}
\end{equation}
In the above relation we assume that the number of permanently nonactive sites
scales as the product of the total number of updates (\ref{7}) with the probability of their creation ($\sim |r|^{1/s}$).
Using (\ref{8}) we obtain that the density of permanently nonactive sites scales as
\begin{equation}
1/l\sim |r|^{1/(2s)}.
\label{9}
\end{equation}
The last relation explains the scaling relation $\bar{e}(r)-\bar{e}(0)\sim |r|^{1/(2s)}$ observed in Monte Carlo simulations of the previous section.
Indeed, for $r$ negative and close to zero the only 'excitations' above the $\bar{e}(0)$
are due to the permanently nonactive sites and this contribution should be 
proportional to their density.

To confirm our scaling arguments we measured the density of inter-chain bonds
satisfying $|w_3|<|-r|^{1/s}$.
For $s>1$ majority of permanently nonactive sites is created on such
bonds.
Our results, for $s=2$ presented in Fig.~\ref{f1}, confirm that the density of
such sites scales as $|r|^{1/(2s)}$.
At the same time this confirms the consistency of the arguments presented in this section.
%%%%%%%%%%%%%%%%%%%%%%%%%%%%%%%%%%%%%%%%%%%%%%%%%%%%%%%%%%%%%%%%%%%%%%%%%
\section{Dynamics as a minimum-finding procedure}
In this section we consider the dynamics of our model as a
procedure to minimalize $e$: active sites are only those whose
contributions to $e$ exceed $-r$.
Of course, the pseudo-energy $e$ is a meaningful quantity also in the active
phase.
Thus, one can ask the following question: In which phase the dynamics finds
more optimal solution.
In the active phase the dynamics cannot inactivate all sites beacause the
requirements for that are too tight.
Thus, even when a 'good' set of sites is found it gets destroyed by
neighbouring active sites.
As a result a finite fraction of sites remains
active with a relatively large contributions to $e$. 
In the absorbing phase a 'good' set has a much larger probability to survive
but the condition for being inactive are milder now and, as a result, worse
solutions are also being found.
Competition of all these effects implies that the $r$-dependence of the
pseudo-energy is rather difficult to predict.

\begin{figure} 
\epsfig{file=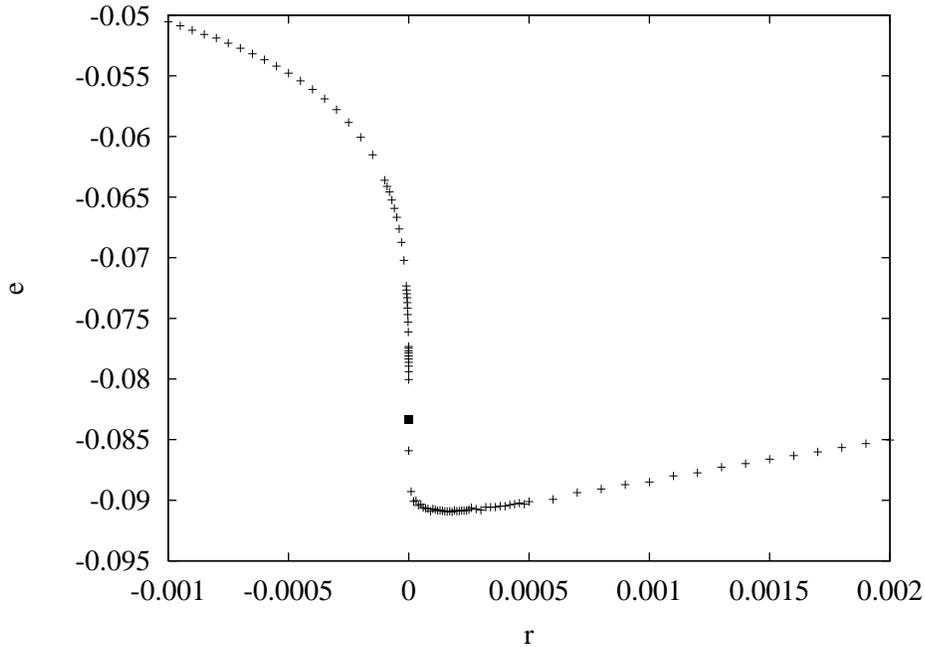, angle=0}
\caption{ The pseudo energy $e$ as a function of $r$
for $s=2$. 
In the absorbing phase ($r<0$) the results are averaged over
$10^3$ absorbing states ($L=10^3$). 
In the active phase ($r>0$) we made an ordinary steady-state averaging
ensuring that the presented results are size independent (for
$r$ close to 0 simulations were made for $L=10^5$). 
The black square denotes the exact value~(\ref{4})
}  
\label{f2}
\end{figure}

To approach this problem we measured $e$ also in the
active phase and our results are shown in Fig.~\ref{f2}.
In the absorbing phase we can see, the already analysed, singular behaviour
with minimum for $r=0$.
More interesting is the behaviour in the active phase.
First, we can see that there is a (narrow) range of $r$ where in the steady
state $e$ is lower than in the absorbing phase.
Most likely, however, for $r\rightarrow 0^+$ the pseudo energy $e$ continuously
approaches the $r=0$ value (\ref{4}).
The lowest-$e$ solution is found for $r\sim 0.00018$ which is very
close to the critical point but not at the critical point.

In the field of combinatorial optimization, one frequently
encounters problems similar to minimization of functions like (\ref{1}).
For example, the traveling salesman problem~\cite{MEZARD} or number
partitioning~\cite{MERTENS} are
equivalent to minimization of certain spin-glass-like Hamiltonians.
Main techniques to deal with such problems are usually simulated
annealing~\cite{KIRK}, genetic algorithms~\cite{BOUNDS} or their hybrids.
In principle it should be possible to extend our approach to solve
some other optimization problems too.
Having an energy function we should define active sites as those whose
contributions to this energy exceed certain value and then evolve the
system similarly to the present model.
Applicability and efficiency of this approach to deal with more typical
optimization problems is, however, left as a future problem.
In addition, we hope that such a method might reveal new connections
between statistical mechanics of models with absorbing states and
computational techniques.
%%%%%%%%%%%%%%%%%%%%%%%%%%%%%%%%%%%%%%%%%%%%%%%%%%%%%%%%%%%%%%%%%%%%%%%%%
\section{Conclusions}
In this work, we have shown that natural absorbing states in a certain
model contain important information about the critical point of the model.
This information is encoded in the static properties of these states.
We should mention that natural absorbing states are well known to contain
information about dynamic properties of models with infinitely many absorbing
states.
In particular, characteristics of the so-called spreading are known to have
power-law singularities at the critical point and the corresponding  exponents
exhibit certain universality.

An extension of this work is to check whether our results are
applicable to other models with infinitely many absorbing states.
Such models appears in various contexts ranging from catalysis~\cite{MENDES} to
self-organized criticality~\cite{VESP} and biological evolution~\cite{LIP2000}.
In some cases, the absorbing states are expected to be quite
complex~\cite{HURTADO} and some indications of criticality might be hidden in
their static structure.
Unfortunately, we do not know which quantities would exhibit these
critical singularities.
In particular it is not obvious to us that in other models there exist
permanently nonactive sites which most likely are responsible for the
singularities observed in our model.
Another question is what are the exponents characterizing these critical
singularities (provided they exist).
In our case, the corresponding exponent ($1/(2s)$) equals to half of the
exponent $\beta(=1/s)$ and it would be interesting to check whether similar
relation hold for other models. 
%%%%%%%%%%%%%%%%%%%%%%%%%%%%%%%%%%%%%%%%%%%%%%%%%%%%%%%%%%%%%%%%%%%%%
%%%%%%%%%%%%%%%%%%%%%%%%%%%%%%%%%%%%%%%%%%%%%%%%%%%%%%%%%%%%%%%%%%%%%%%%%
\acknowledgements
This work was partially supported by the Swiss National Science Foundation.
%%%%%%%%%%%%%%%%%%%%%%%%%%%%%%%%%%%%%%%%%%%%%%%%%%%%%%%%%%%%%%%%%%%%%%%%%%

%%%%%%%%%%%%%%%%%%%%%%%%%%%%%%%%%%%%%%%%%%%%%%%%%%%%%%%%%%%%%%%%%%%%%%

\begin{references}
\bibitem{COMM1} In~\cite{LIP2001} the variables are from the interval
(-0.5,0.5) but this difference does not affect the critical behaviour of the
model.
\end{references}
\end{document}